# Photoassisted physical vapor epitaxial growth of semiconductors: a review of light-induced modifications to growth processes

Kirstin Alberi[1] and Michael A. Scarpulla[2,3]

[1] National Renewable Energy Laboratory, Golden, CO 80401
[2] Materials Science & Engineering, University of Utah, Salt Lake City, UT 84112
[3] Electrical & Computer Engineering, University of Utah, Salt Lake City, UT 84112

**Abstract**

Herein we review the remarkable range of modifications to materials properties associated with photoexcitation of the growth surface during physical vapor epitaxy of semiconductors. We concentrate on mechanisms producing measureable, utilizable changes in crystalline perfection, phase, composition, doping, and defect distribution. We outline the relevant physics of different mechanisms, concentrating on those yielding effects orthogonal to the primary growth variables of temperature and atomic or molecular fluxes and document the phenomenological effects reported. Based on experimental observations from a range of semiconductor systems and growth conditions, the primary effects include enhanced anion desorption, molecular dissociation, increased doping efficiency, modification to defect populations and improvements to the crystalline quality of epilayers grown at low temperatures. Future research directions and technological applications are also discussed.

**Introduction**

Advances in electronic device technologies are often enabled by the ability to synthesize semiconductor materials with improved properties. Crystal growth must provide precise control of the material composition and purity, structure, defect concentrations and interfaces in order to allow desired properties, high performance devices, and new physical phenomena to manifest. Vapor phase epitaxy (VPE) methods, in which the constituent elements are supplied via gaseous precursors or evaporated atoms or molecules, has been utilized to great advantage for device fabrication because they permit high-purity layer-by-layer growth of single crystalline material with this control and flexibility. Variants of molecular beam epitaxy (MBE) and organometallic vapor phase epitaxy (OMVPE / MOCVD) are among the most widely used VPE techniques.

One advantage of VPE techniques is that they can operate over a range of energetically- and kinetically-limited regimes, allowing for growth of desired materials that may be strictly thermodynamically unstable.[1,2] For example, MBE and OMVPE can be used to grow thin layers of alloy compositions and even phases not stable in bulk crystal form because they can exploit epitaxial strain, surface energies and reconstructions, and diffusion kinetics that are different from the bulk. However, the degrees of freedom offered by VPE alone are not great enough to overcome all the semiconductor synthesis challenges. Exciting the growth interface with light or other "hot" (non-thermalized) energetic particles presents an option for increasing the range of control during epitaxy through additional changes induced at the growth surface. In this article we review concepts for modifying semiconductor VPE using directed energy, especially in the form of photon irradiation.

Despite the generally excellent material quality offered by MBE or OMVPE, it is not always possible to realize semiconductor layers with the desired combination of composition, doping, structure, or crystalline and optoelectronic quality. In general, growth at high temperatures promotes adatom diffusion on the growth surface and is typically best for achieving high-quality material. However, enhanced surface migration can lead to composition segregation in cases where solubility limits are exceeded. Furthermore, atoms may migrate within epitaxial layers, aided by vacancies or electric fields introduced by Fermi level pinning at the growth surface. The resulting diffuse doping or alloying profiles can compromise device performance. Especially in cases of extreme mismatch of ionic radii, alloy compositions desirable in terms of bandgap or lattice constant cannot be achieved either because of segregation and spinodal decomposition or pileup at the growth surface without incorporation. Native defect populations are also notoriously difficult to control in doped semiconductors. Low or high values of the Fermi level from extrinsic doping reduces the formation energy of compensating defects. The effect ultimately imposes doping limits, most starkly for wide-bandgap semiconductors in which the Fermi level and thus defect formation energy can shift over large energy ranges. The last two processes, dopant diffusion and defect formation, are directly linked to the Fermi level thus coupling them to the extrinsic doping concentration. These challenges may be partially overcome by reducing the substrate temperature to allow the chemical potentials imposed by the incident fluxes to more heavily influence the growth processes. For example, in a limiting case of very low substrate temperature, no desorption and no diffusion on the growth surface could occur and thus the grown composition would be uniform and identical to the ratio of arriving fluxes. Yet this solution comes at the expense of crystalline quality. The example above would yield amorphous growth at the same time. Thus the combination of crystal quality, composition, and properties of epilayers grown by VPE can be conceptualized as the result of a fine balance adatom arrival, desorption, diffusion and incorporation.

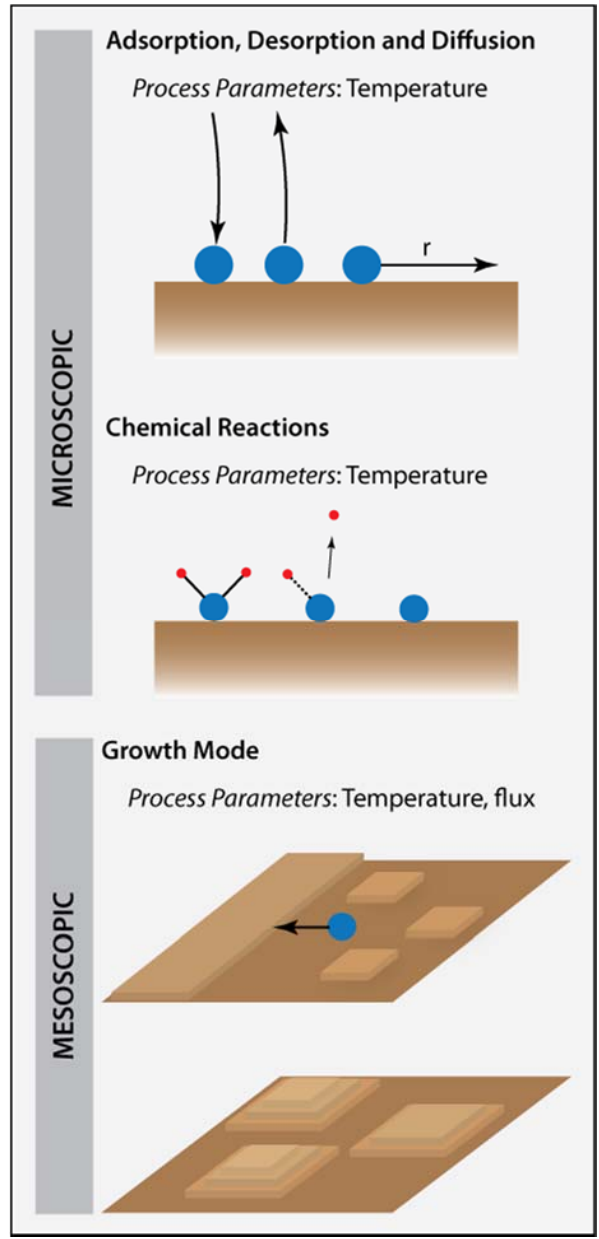

**Fig. 1** Elementary processes involved in semiconductor epitaxy

The representative trade-offs discussed above arise from the fact that this balance is only mediated by a few externally controlled parameters: temperature and the atomic or molecular fluxes arriving at the growth surface. The chemical identity of the species arriving at the interface, their equivalent partial pressures based on the impingement rate, and their states of excitation determine the chemical potential. The relative rates of adatom "sticking" and thermally-excited desorption determines the crystal growth rate. These few control parameters provide limited means for manipulating the various microscopic growth processes independently of one another (see Figure 1).

Another parameter that may yield additional degrees of freedom is non-thermalized energy, which can be provided to the growth surface in the form of any energetic particle – photons, electrons or accelerated ions. Since the critical phenomena of epitaxial growth occur primarily at the growth surface, the most selective stimulation processes will deliver energy only to the surface. In most cases, energy is absorbed over greater depths – for example high energy radiation or longer wavelength but still above bandgap light. Combinations of surface and bulk effects will result in these cases.

Ion beam assisted deposition (IBAD) was explored in the late 1990s as a way to increase the effective adatom mobilities, especially in the context of polycrystalline films but in epitaxial layers as well[3,4]. Low energy ions can transfer large momentum to adatoms, causing them to redistribute over greater distances as well as preferentially sputtering weakly bonded and low-coordinated atoms in the topmost layer. The first-order objective of employing IBAD is therefore to aid the migration of poorly-bonded atoms and achieve a higher degree of crystalline perfection. However, bombardment of the growth surface with energetic ions can modify the stoichiometry and surface reconstruction, change reaction rates, and alter growth kinetics. Improved surface smoothness, enhanced adatom mobility and interdiffusion, a transition from 3D to 2D growth modes, and preferential etching of incipient phases have been demonstrated.[3,4,5] Drawbacks to IBAD include structural damage, vacancy formation, amorphization and grain size modification.

Alternatively, photon irradiation of the growth surface has the potential to address several of the growth challenges noted above. Absorption of light can affect processes at the growth surface through three primary pathways: photo-dissociation of molecular precursors, local enhancement of thermal energy, and generation of free carriers, which influence bonding processes, enhance desorption and move the positions of the quasi-Fermi levels (QFLs). The absorbed energy and photogenerated carriers can therefore alter a wide range of thermodynamic and kinetic processes. This last pathway, in particular, offers an additional parameter to decouple growth processes influenced by the Fermi level form the extrinsic doping conditions. Illumination is an attractive means of exciting the growth surface, as photons impart energy with negligible momentum compared to ions. Lastly, irradiation can be made highly uniform, and the selectivity of physical processes excited can be achieved through the manipulation of the incident spectrum.

One of the earliest reports of using light to improve semiconductor crystal growth was in 1961 by Kumagawa and Nishizawa.[6] It was followed by two more in-depth studies in 1968: one by the same authors[7] and another by Frieser.[8] Both groups demonstrated that light from a mercury lamp could reduce the substrate temperature needed to achieve Si epitaxy by chemical vapor deposition (CVD). In the 1980s, a number of research groups began to apply photo-assisted epitaxy techniques to the growth of Si, III-V and especially II-VI semiconductors by OMVPE and MBE. Irradiation of the growth surface by above-bandgap photons allowed II-VI epitaxy to proceed at lower substrate temperatures, and it was also found to improve both crystalline quality and the doping efficiency in CdTe, HgCdTe, ZnSe and ZnS.[9,10,11,12,13] Much of this work was curtailed around the same time that the intense II-VI semiconductor research of the 1980s and 1990s tapered off in favor of wide bandgap nitride development. Yet recently there has been renewed interest in photoassisted VPE as a mean to overcome new challenges associated with the synthesis of increasingly complex materials, wide bandgap semiconductors and heterostructures.[14,15,16]

The review is organized as follows: we first provide an overview of epitaxial growth processes and subdivide physical mechanisms in terms of the type of photoabsorption. We then discuss observed phenomena and associated hypothesized mechanisms reported in the literature. We conclude by identifying areas for future research and development. The insights compiled here are meant to inform new uses for this technique. It should be noted that the discussion is strictly limited to techniques where photons are directed onto the semiconductor growth surface, as opposed to techniques utilizing light to ablate source material (e.g. pulsed laser deposition PLD or laser MBE).[17]

**Introduction to Vapor Phase Epitaxy and Growth Mechanisms**

Vapor phase epitaxy techniques, including MBE and OMVPE, are widely used for semiconductor growth because they offer extraordinary control over the growth rate, chemical potential and material purity. Combined with specific choice of substrate temperature and material, these factors allow the control of adatom incorporation dynamics, crystalline phase, doping, and interface and defect formation. We start with a brief review of some of the basic growth mechanisms that will allow us to better evaluate how light can affect growth processes and the resulting material properties. More in-depth information about semiconductor epitaxy and growth mechanisms, particularly as they pertain to specific growth techniques, can be found in Refs 1-3, 18 and 19. Some of these processes are depicted in Fig. 1.

Vapor phase epitaxial crystal growth involves the incorporation of impinging atoms from a gas phase into a solid crystalline structure and occurs when the rate of atomic incorporation into the crystal exceeds the rate of desorption. Generally, the incorporation process proceeds as impinging atoms or molecules physisorb and/or chemisorb onto the substrate and diffuse across the surface until they encounter a suitable site for incorporation via the formation of multiple chemical bonds.  In OMVPE, a form of cold-wall chemical vapor deposition, molecules are introduced temperatures lower than the substrate and arrive at the surface via diffusion through the flow boundary (or stagnation) layer near the substrate.  Thus at most their impinging temperature is that of the substrate and thus kinetic energies are small fractions of eV.  In MBE, species are delivered in the molecular flow or ballistic regimes from Knudsen cells held at elevated temperatures relative to the substrate.  Although source temperatures may exceed 1500 °C, their kinetic energies are still small fractions of an eV. The ability of an atom or molecule to initially "stick" to the surface requires that it transfers enough energy to the substrate so that it cannot easily desorb at very short timescales. This process therefore requires the substrate temperature to be much lower than the temperature of the impinging atom or molecule. Attachment of atoms through chemisorption, where the adsorbate undergoes a chemical reaction with the substrate, is dependent on the local bonding environment on the surface. Physisorption, where the electronic structure of the adsorbed atom or molecule is perturbed (i.e. through the formation of van der Waals bonds), is often less dependent on the surface geometry or environment. Both processes may be accompanied by dissociative chemical reactions that break apart molecules, allowing eventual chemisorption.

The maximum distance an adatom can potentially diffuse laterally on the (approximately) two-dimensional (2D) growth surface is $r_m = \sqrt{D(T)t_{res}}$, where $D(T)$ is the 2D surface diffusion coefficient and $t_{res}$ is the residence time in the chemisorbed state (i.e. before either desorbing or incorporating with a higher degree of bond coordination). However, the mean distance an atom diffuses, $\langle r \rangle = \sqrt{D(T)t_{ave}}$ is determined by the average time, $t_{ave}$, allowed for diffusion. Since the atomic or molecular flux, $J$, in part determines how frequently adatoms encounter one another and (in competition with desorption rate) how fast each layer is deposited, $t_{ave} \propto 1/J$.[19] Ideally, $J$ should be low enough such that adatoms have sufficient time to move around on the substrate surface and find the lowest energy incorporation sites. Excessive fluxes can shorten that time and prevent adatoms from diffusing far enough to find the most favorable incorporation sites. The diffusion coefficient characterizes the jump frequency of the atom with an activation energy $E_a$. While the local environment on the surface can influence $D(T)$ (for example, anisotropic diffusion along dimer rows or diffusion barriers over step edges), the activation energy typically gives it an Arrhenius dependence.

Desorption or evaporation is also governed by an atom's attempt to escape frequency which can be estimated from a harmonic oscillator model given the mass and bonding potential. This can be affected by its local bonding environment as well as any chemical reactions that must take place to aid desorption (for example, the formation of $As_4$ molecules from two $As_2$ species on the surface). In most cases, though, the desorption rate can also be approximated with an Arrhenius dependence.

Chemical reactions may need to occur in order to produce atomic species that can then incorporate into the growing crystal. Some examples include dissociation of metal-organic and hydride precursor in the case of CVD-based growth or of $As_4$ into $As_2$ dimers and single As atoms in the case of MBE. Because these chemical reactions have an associated activation energy, substrate temperature is often the main driving force for molecular dissociation – commonly called pyrolysis in the OMVPE literature. A key consideration is to design the reaction to have an activation energy that is large enough so that it doesn't proceed on the walls of the growth chamber or in the gas itself but is low enough so that it can be promoted at reasonable substrate temperatures. Larger molecules of anion species produced by evaporation from a solid source during MBE growth must also dissociate as part of the growth process. This may include $As_4$ or $Se_n$ (n > 2) molecules. Atoms can leave the growth surface as molecules as well. Both of these processes can influence growth rates. Beyond the substrate temperature, other factors that can affect the dissociation rate include substrate orientation, the presence of specific sites on the lattice that promote or catalyze reactions, surfactant atoms or component species monolayers on the growth surface, and the presence of free carriers that may transfer to the arriving species and help to break bonds at the growth suface.

On a mesoscopic level, epitaxy is generally divided into layer-by-layer 2D or three dimensional (3D) island growth modes. The growth mode is determined by a number of factors, including the atomic or molecular flux, the range of surface species migration and the surface structure of the substrate. The 2D layer-by-layer formation process more often results in layers of high optoelectronic quality. On a planar surface with no defects, adatoms randomly form small nuclei that become stable clusters if they reach a critical size. Those clusters then coalesce into a full layer, and the process repeats for the next layer. A terraced surface resulting from vicinal off-cut from a low-index crystal plane offers ready-made incorporation sites with higher degrees of bonding coordination along the terrace edges. This allows the growth to proceed in a step-flow manner, provided that the atomic diffusion length is long enough for adatoms to traverse the length of the terrace. Local changes in the incorporation rate, the geometry of incorporation sites or surface diffusion length can lead to enhanced surface roughening. Growth can be forced into a 3D mode when many layers begin to grow at once, creating a multi-tiered surface. This can happen when *J* is high, <*r*> is low, or energetic barriers to adatom movement over step edges (so-called Schwoebel barriers) are high, preventing stable clusters from coalescing before adatoms start to form a second layer. Growth on a substrate with significant topology can also perpetuate 3D growth modes.

Based on this abridged description of growth mechanisms, a discussion of thermodynamically vs kinetically limited growth regimes is warranted. While true thermodynamic equilibrium is incompatible with crystal growth – thermodynamic equilibrium means the system is static (i.e. no growth) and there are no net fluxes of heat or mass across the system boundary – epitaxial growth can occur relatively close to or far from thermodynamic equilibrium. The adatoms, islands of atoms, and steps between atomic layers can be considered as separate gas, liquid, and solid phases confined to the 2D surface that mediate between the 3D vapor, liquid, and solid phases to which equilibrium phase diagrams typically refer. Local equilibrium at the growth interface means that the various exchange rates of mass and energy between these bulk and surface phases satisfy conditions of detailed balance. This defines the microscopic statistical mechanics dynamic equilibrium, which is equivalent to the thermodynamic concept of reversible equilibrium.

High-quality VPE usually occurs close to local equilibrium, where the condensation rate from the vapor only slightly exceeds the corresponding desorption rates and the adatom diffusion length is sufficient to enable sampling of a large number of available sites. Under these conditions, the growth interface achieves local equilibrium and steady state with the strains, bonding configurations, perimeter/edge energies, surface energies, arrival and desorption rates, and electrochemical potential imposed by the substrate and growth parameters.[20] The exception occurs at high arrival fluxes and low substrate temperatures when even local equilibrium is broken because the times required for desorption or diffusion to an optimal incorporation site are much longer than the time between arriving atoms.

MBE, OMVPE, and related techniques may be also used to grow crystals that are not thermodynamically stable in bulk form. One option is to draw upon the difference in the energetics of the 2D surface compared to the 3D bulk. The lateral adatom diffusion length <r> on the 2D surface is usually much greater than the 3D bulk diffusion length $r_{3D} = \sqrt{D_{3D}(T) t_{growth}}$ (where $D_{3D}$ is the bulk diffusivity and $t_{growth}$ is the time of the epilayer). Negligible 3D bulk diffusion therefore allows metastable configurations and abrupt interfaces to be frozen into the epilayer, especially at lower growth temperatures and high chemical potentials of constituent atoms supplied by the vapor phase.

This conceptual framework allows us to better envision the effects of applying non-thermalized excitation (i.e. light with a spectrum differing from the blackbody emission associated with the interface temperature or accelerated ions or electrons) to the growth surface. Generally, it will change the kinetic rates of various growth processes and thus result in a new dynamic equilibrium. These changes can range from negligibly small to quite dramatic. It may also produce trivial or non-trivial effects on the grown crystals in terms of available control variables and growth degrees of freedom. Here, the term "trivial" implies that the end state of the grown crystal can be reproduced within the existing phase space permitted by a combination of substrate temperature and atomic/molecular arriving fluxes. The addition of the non-thermalized energy does not add an independent control variable to the growth process. "Non-trivial" effects are those that enable unique material and sample property end states, or achievement of the same end state while also optimizing an additional variable. Examples of unique end states would be enhanced doping density and/or minority carrier lifetime compared to non-stimulated growth. Examples of the latter type of non-trivial effect would be shorter growth time with material properties constant or enhanced structural quality within the same growth time. The non-trivial effects are of more practical interest as they enable the possibility of achieving otherwise inaccessible combinations of growth process characteristics and resulting crystal properties. Specific mechanisms of modification are outlined below.

**Categorization of Photoinduced Effects on Vapor Phase Epitaxial Growth**

From the above discussion, it is evident that absorption of light or other forms of excitation can influence processes at the growth surface through supply of excess thermal energy and photogenerated carriers. We can classify the resulting effects into four categories according to the location of the energy absorption: 1) photochemical effects in vapor phase, 2) localized surface heating or photochemistry of adsorbed species, 3) sub-bandgap photoabsorption involving defect states in the semiconductor, and 4) bulk band to band absorption resulting in excess carrier effects. We provide an overview of these potential effects briefly here before discussing them further in the context of specific experimental and theoretical investigations. A schematic of these categories is presented in Fig. 2.

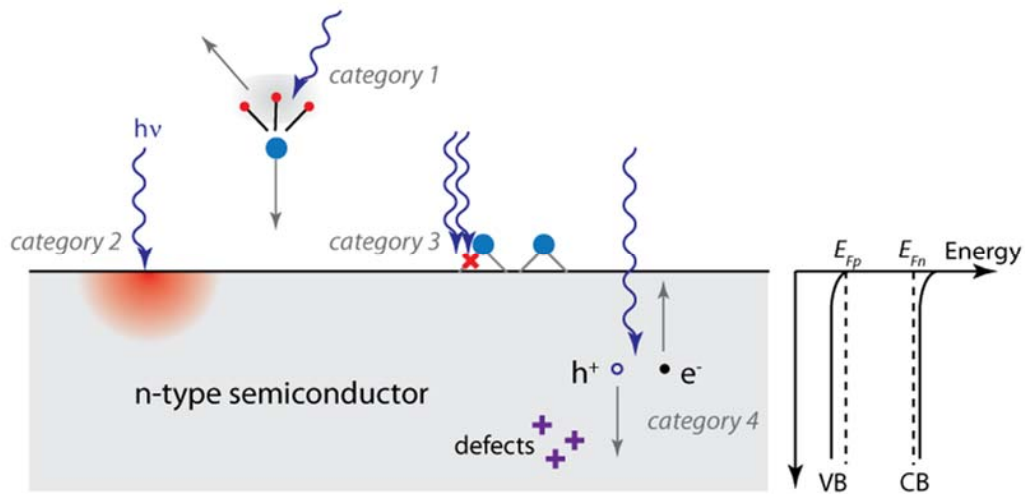

**Fig. 2** Schematic of potential photo-induced effects from categories 1-4 in the text.

*Category 1 - Photochemical Effects in Vapor Phase*

Examples of photochemical effects in the 3D vapor phase include excitation and photo-dissociation of gas phase precursors, atoms, or clusters. The terms photo-induced and photo-assisted epitaxy by in large refer (respectively) to OMVPE growth driven entirely by light or using it to assist the typical thermal excitation.[21,22,23,24] The probability of photochemical effects in the gas phase will be proportional to the molecular absorption cross section, the pressure of the reactant, and the photon flux. Since OMVPE typically occurs at higher pressures in the $10^{-3}$ to $10^3$ Torr range compared to $10^{-9} - 10^{-5}$ Torr beam equivalent pressure for MBE, such reactions in OMVPE are many orders of magnitude more probable. Results for other chemical epitaxy processes (e.g. hydride vapor phase epitaxy HVPE) should be similar. We note however that the heating method used in different OMVPE reactors – wide-spectrum incandescent lamps versus inductive heating – may produce different results under nominally the same temperature, pressure, and flow rates for such reasons. Because this class of effects has been sufficiently reviewed in the context of OMVPE, we abridge further coverage herein.

In the case of physical vapor epitaxy such as MBE, the extremely low density of atomic/molecular beams (of order $10^{11}$ atoms/cm$^3$ even for the $10^{-5}$ Torr beam-equivalent pressures for anion species), combined with practically achievable laboratory photon fluxes make the probability of such events extremely small. The influence of light on the dissociation of molecules in the vapor phase is therefore expected to be quantitatively insignificant. In some sense though, the use of Knudsen cells with high temperature crackers, heated surfaces in the case of hot-wire chemical vapor deposition, or plasma excitation achieve the same goals – increasing the chemical potential of the desired species in the gas phase compared to the molecule or atom in its ground state by exciting internal degrees of freedom or dissociation.

*Category 2 - Localized Surface Heating or Photochemistry of Adsorbed Species*

In this category, we consider that both energy absorption and its effects on growth processes are limited to the growth surface – i.e. there is no transport of carriers or heat to the growth surface along the growth direction. The influence of heat and carriers differ in the mechanism and thus possible absorption spectra. Localized surface heating would arise from above-gap excitation of the semiconductor but with short wavelengths and short pulse times localizing heat to the growth interface. Photochemistry of adsorbed species, for example photodissociation of uncracked $As_4$ tetramers, would be driven by absorption into the localized states of those species and thus may be able to utilize sub-bandgap wavelengths. We speculate that this mechanism could be used to grow very unstable compositions at higher crystalline perfection by keeping the grown layer cold.

In principle during steady-state growth, acceleration of the transition and transport processes at the 2D growth surface without corresponding 3D processes can be achieved by differential heating of the 2D growth interface. However, in all but the very most extreme cases such effects will be negligible for two reasons. First, it is nearly impossible to achieve a sufficiently large thermal difference through a crystalline substrate (note that the substrate and not the epilayer provides the lager thermal resistivity based on its much greater thickness). Second, the effects will usually be trivial – that is achievable also by simply raising both the bulk and surface temperatures. The first point can be illustrated quantitatively. At high temperatures, most semiconductors ranging from Si to InSb and CdTe have thermal conductivity of order 0.1 W/cm$^2$. Alloyed semiconductors indeed have lower conductivity but are rarely available as substrates for epitaxy. Assuming heating from the back and only perfect blackbody emission from the front, the surface and backside temperatures can be computed for imposed heat fluxes. For surface temperatures in the 100-1000 °C range, the temperature difference from front to back will be less than 10 °C and through a 1 μm epilayer will be a small fraction of a °C. Another limiting case of heating the growth surface in steady state only by light and actively cooling the back surface can achieve a 10-100 °C temperature difference in a typical semiconductor wafer (and of course much smaller through the epilayer) only for impinging light fluxes of order 1000 W/cm². Thus the possibility of such effects being observable during steady-state growth is fairly remote.

On the other hand, the use of a repetitive pulsed energy source such as a pulsed laser can induce transient heating of the surface and near-surface region to values even above the melting temperature without appreciably heating the substrate. This is because of the massive instantaneous flux but low fluence in the pulses, easily in the $10^8$ W/cm² range for <1 J/cm² for 10 ns lasers. For ns lasers the fluence range below approximately 1-100 mJ/cm² corresponds to solid phase transient annealing. The next highest range of approximately 0.1-1 J/cm² corresponds to pulsed laser melting (PLM), which is a remarkably rapid (few m/s growth rate) form of liquid phase epitaxy. Finally, fluences >1 J/cm² correspond to laser ablation.[25,26,27,28,29,30,31,32,33,34] The fluence ranges mentioned above are for approximately 10-30 ns pulses and scale as t^½ where t is the pulse duration. However, in the short ps and fs regimes the carriers and lattice do not thermalize within the pulse so other physics begins to apply.

If the carriers and lattice thermalize, the characteristic distance heated is given by the larger of the optical absorption depth $=1/\alpha(\lambda)$ or the heat diffusion length $=\sqrt{D_{th}\tau_{pulse}}$ wherein $\alpha(\lambda)$ is the wavelength-dependent absorption coefficient (cm⁻¹), $D_{th}$ is the thermal diffusivity (cm²/s) and $\tau_{pulse}$ is the pulse duration (s). Given typical thermal diffusivities for semiconductors, pulses must be in the fs to approximately few-ns range in order for the optical absorption coefficient to set the characteristic depth. Otherwise, the heated depth will be in the few-hundred nm range for typical ns lasers with 5-50 ns pulses, independent of wavelength. The limiting case resulting in the maximum differential heating would combine a low-thermal conductivity epilayer of an alloy or heavy compound, photon energy sufficiently above bandgap to achieve $\alpha=10^6$ cm$^{-1}$ for <10 nm absorption depth, and ps-fs pulses. The heat flow time will typically be on the scale of ns in the fluence range of interest, well below the surface melting threshold. Thus, the repetition rates of ultrafast lasers in the 10-100 MHz range will still separate the effects of each pulse sufficiently that their effects will not overlap in time. Higher duty cycles would increase the magnitude of observable effects. In this case, a purely thermal mechanism of differential heating of the growth surface may be exploitable for non-trivial effects because of exponential dependences of thermally activated processes and the typically lower activation energies of surface processes.[35,36,37] To our knowledge, experimental proof of such a mechanism has not been reported, making this an area for further investigation.

*Category 3 - Sub-bandgap Photoabsorption Involving Defect States in the Semiconductor*

Photon absorption can generate excited states at the surface, including broken or reconfigured bonds and various charge states. These excited states may influence the chemical reactivity of the surface. Sub-gap absorption into transitions between defect and band states can also change the concentrations of free carriers. In both of these cases, the total number of available transitions per sample area are much smaller than for bulk absorption and so such effects are probably much less significant than those involving excess carriers transported towards the surface after bulk absorption.

In some cases, adatom diffusion lengths are reduced because adatoms interact with the excited surface states and form bonds and thus their probability of jumps to adjacent sites is suppressed. In others, surface species may be dislodged and even desorbed from wrong-bonded sites as well as low-coordination surface sites such as isolated adatoms, edge sites, and kink sites compared to those incorporated at a step or island. This would tend to decrease growth rate yet suppress the incorporation of bonding defects (e.g. vacancies and antisites). Another class of effects in this category would be changes in the charge states of surface atoms, which may affect the final alloy composition, incorporation of extrinsic dopants, and the concentrations of native defects. Lastly, there is solid evidence that 3D diffusion can be significantly modified by photoabsorption. Specifically, photoabsorption can generate excess carriers, heat the growth surface by local multi-phonon release or modify of the diffusing atom's charge state and thus the energy landscape and activation energy for diffusion.[38,39,40,41, 42, 43,44,45,46] We are unaware of direct experimental results for 2D surface diffusion enhanced via absorption of above-bandgap photons specifically by surface states. However, such effect may be possible.

Absorption involving band-to-defect transitions (photoionization or photoemission) in the bulk are at least 100 or more times less-probable than band-to-band transitions but can change the charge state of these defects and thus the local charge balance and space charge region. This in turn can affect the formation enthalpy for point defects as well as affecting the incorporation of dopants and native defects during growth.

*Category 4 - Bulk Band to Band Absorption Resulting in Excess Carrier Effects*

The charge states of pre-existing point defects and surface atoms can be modified by excess carriers generated by bulk absorption. These mechanisms can affect both the growth surface and the already-grown epilayer, which is essentially annealed for the duration of growth. For example, photogenerated holes swept to the surface by the electric field induced by surface Fermi level pinning in n-type material can weaken or break dimer bonds in the surface reconstruction, altering the incorporation of dopants or alloying atoms at the growth surface. In this example, the mechanism is similar to the effects of surfactants.[47,48] Once carriers arrive at the surface, they can weaken bonds and increase desorption rates – for example changing CdTe stoichiometry to Cd-rich while holding the fluxes and temperature constant. Enhanced desorption can also introduce vacancies, which may persist in the growing layer. Related phenomena have been observed in CdTe, Si, and $TiO_2$ at surfaces upon above-bandgap illumination. ,[38,39,49] These vacancies can also be filled by dopant or alloying species, increasing the effectiveness of doping or alloying under otherwise unchanged growth conditions.

Additionally, photoexcitation can produce non-equilibrium concentrations of electrons and holes, which can alter defect state populations directly through the energetics of defect charge states.[14,50] Excess carriers can also screen the electric field near the surface and modify diffusion, both of which will affect the distribution of dopants and defects throughout the grown layer.[38,39,40] Such mechanisms have been known for decades, although to our knowledge the competing theoretical explanations have not been resolved.

Lastly, point defects can be directly created in the bulk by the energy released from band-to-band recombination. The earliest example of this was color center formation in I-VII compounds, which are characterized by weak, ionic bonding[51]. Such centers are charge-neutral native defect complexes consisting of Schottky or Frenkel pairs produced by the energy released from photocarrier recombination. In order for this type of photogeneration of defects to occur, the first criterion is that the bandgap is larger than the energy required to produce the defect (e.g. move an atom off-site into an interstitial in the case of Frenkel defect pairs). The second criterion is that the carrier-lattice coupling is strong.[43] Detailed mechanisms are still debated and may be different in different cases. There exists some evidence of such carrier-induced defect formation extending into the slightly-more-covalent II-VI compounds such as CdTe.[52, 53, 54] Also, AX defect formation (AX being the acceptor analog of DX centers) in CdTe has been predicted computationally and recently observed.[55,56,57] This process of Schottky or Frenkel pair formation upon carrier capture can be considered as a more pronounced configuration change into adjacent lattice positions compared to the relaxations accompanying transitions to the deeper state in DX centers for III-V and II-VI materials.[58] The metastability of AX or DX centers in more covalent compounds compared to the more irreversible (thus non-metastable) Frenkel or Schottky pair configurations in ionic compounds is related to the energy barrier for switching between the on-site and distorted or off-site configurations. Because of the changes in configuration accompanying switching between the multiple states of DX or AX centers or the presence versus absence of carrier-induced point defect pairs, the thermal activation barrier and optical transition energy required to switch between these states may not be equal.

Overall, our analysis of the likelihood of the effects in the first three categories occurring is relatively low unless very high energetic fluxes are applied. We conclude that category 4 – bulk band to band absorption – is the type of absorption that can produce the most dramatic and useful changes to growth for typical physical vapor epitaxy carried out in steady state.

**Experimental Reports of Photoassisted Epitaxy**

Having identified pathways through which photon irradiation can affect growth processes, we review the growth modifications that have been reported in the literature. As noted above, excessively high photon fluxes can of course lead to an increase in the substrate temperature. However, under controlled illumination conditions, most growth modifications have been demonstrated to be primarily electronic, rather than thermal, in nature. The following sub-sections detail specific mechanisms of photo-assisted epitaxy that have been observed and investigated to date. Where possible, we have identified material trends. In evaluating these results, it is important to keep in mind that more than one effect may be present and that changes can occur to varying degrees depending on the photon irradiation conditions as well as the other growth parameters.

*Adatom Desorption and Reduced Growth Rates*

Enhanced desorption of surface species was among the first studied effects introduced by photon irradiation of the growth surface. The change in the desorption constant of cation and anion species was directly measured by Benson, *et al.*, for the case of Cd, Sb and Te desorption from a CdTe surface during MBE growth.[11] Without photon irradiation, Te desorbed with an activation energy of 1.9 eV, while Cd desorbed with an activation energy of 5.1 eV (see Fig. 3a). At substrate temperatures below 300 C, irradiation by a 26 mW/cm$^2$ He-Ne laser lowered the Te desorption activation energy to 0.15 eV, while irradiation by a 75 mW/cm$^2$ Ar ion laser led to a slight further reduction. The Cd desorption activation energy was reduced by a smaller amount to 4.3 eV, and there was even less of an effect on the Sb dopant atoms. Selenium atoms were also found to have a higher desorption rate than Zn from a ZnSe surface under photon irradiation.[59] These results suggest that photon irradiation affects the host anion desorption much more than cation desorption. In fact, high photon fluxes have been found to severely reduce II-VI semiconductor growth rates or arrest growth altogether.[10,60]

Enhanced desorption rates have been linked to modification of the adatom charge states by photogenerated free carriers. The prevailing explanation is that photogenerated carriers provide extra charge to chemisorbed species that allow them to desorb either as neutral atoms or as molecules, aided by thermal energy, as depicted schematically in Fig. 4.[59,61,62] Consider a compound semiconductor AB. Hole transfer to negatively charged anions, B$^-$, bonded to underlying cations weakens the chemical bond and creates B$^0$ species. The desorption flux of the B$_2$ molecule, $J_D$, then rises due to the increased concentration of B$^0$ atoms:

$$J_D(B_2) = K_D \left[ B^0 \right]^2 \qquad (1)$$

where $K_D$ is a constant.[61] A similar process can occur for the cation through the capture of photogenerated electrons. Physisorbed species can also capture free carriers due their ionicity. Photogenerated electrons are transferred to the anion species, while holes are transferred to the cations. Consequently, the species may less readily chemisorb to the surface and instead desorb at higher rates.[59] The elevated desorption rates of anions under photon irradiation may be partially due to their already lower activation energy compared to cations. Furthermore, the II-VI semiconductors for which this effect has been most noted are almost all n-type, even when not intentionally doped. Fermi level pinning at mid-gap surface states will set up slight band bending in that case, which will drive holes toward the surface, promoting desorption of chemisorbed anions. Similar effects have been observed under electron irradiation.[59]

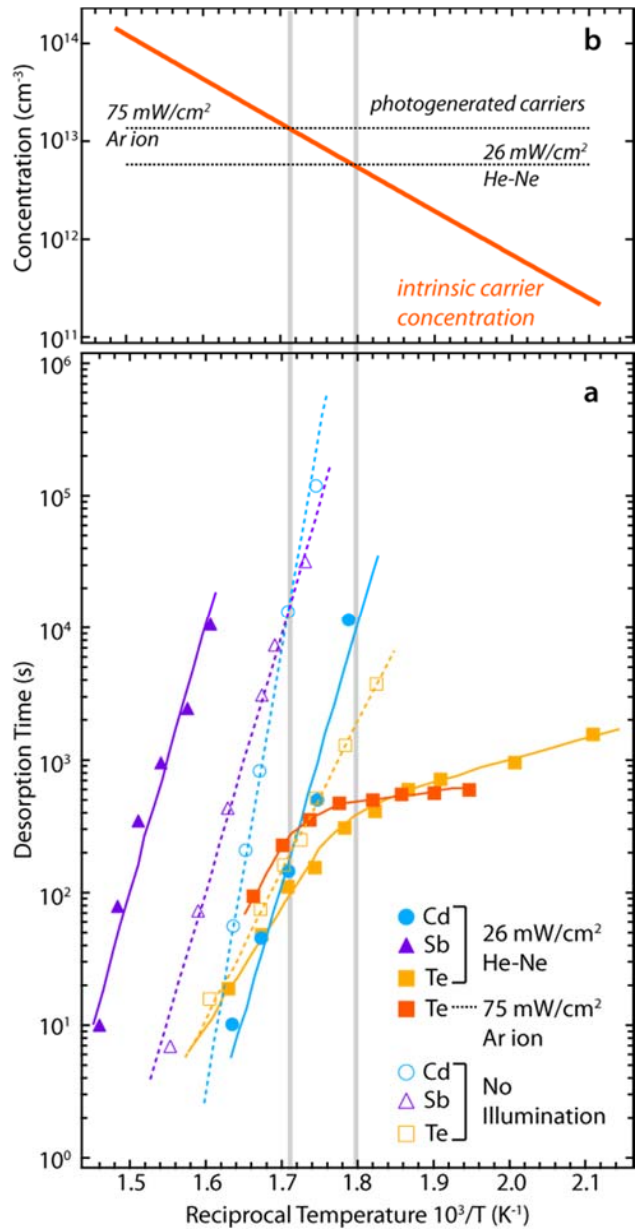

**Fig. 3** Desorption of species from a CdTe surface. (bottom) Desorption time as a function of substrate temperature for Cd, Te and Sb atoms as a function of illumination conditions. (top) Intrinsic carrier concentration in CdTe as a function of substrate temperature. The photogenerated carrier concentration for two laser illumination conditions are also marked. Experimental data is from references 11 and 63.

To investigate the possible role of photogenerated carriers further, we calculated the intrinsic carrier concentration of CdTe over the same reciprocal temperature measured in the desorption study by Benson *et al.,* and have plotted it in Fig. 3b. We also plotted the carrier concentrations expected to be generated by the three illumination conditions used in Refs 11 and 63, assuming a carrier lifetime of 1 ns and an absorption depth of ~140 nm. The temperatures at which the Te desorption activation energy abruptly decreases roughly correspond to the temperatures at which the photogenerated carrier concentrations start to exceed the intrinsic carrier concentration. This rough estimate further supports the hypothesis that photogenerated carriers play a role in enhancing the anion desorption rate and suggests that their concentration must exceed the intrinsic carrier concentration to have a measurable effect. We should note that another study also looked at changes in the Te desorption from a CdTe surface.[64] A reduction in the desorption time upon the addition of illumination was observed but not a change in the activation energy. Differences in photon irradiation conditions may be responsible for the different outcome.

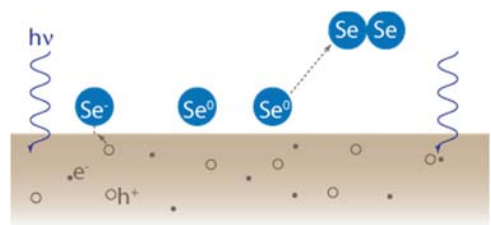

**Fig. 4** Schematic of Se desorption mediated by photogenerated holes.

*Molecular Decomposition and Increased Growth Rates*

In the case of growth using molecular precursors, photon irradiation can play a role in molecular decomposition and alter growth kinetics. Elevated substrate temperatures are typically required to dissociate oganometallic and hydride molecules used in CVD-based growth methods, which constrains control of other kinetic and thermodynamic processes. Photons can alternatively be used to drive molecular decomposition through two pathways: gas phase dissociation and dissociation of adsorbed molecules on the substrate surface. In both cases, photon irradiation lowers the thermal requirements for dissociation, allowing growth to be carried out at lower substrate temperatures.

Gas phase dissociation involves photon absorption directly by the molecule before adsorbing onto the substrate surface. Experiments designed to test this mechanism are often carried out by directing the light source parallel to the substrate surface to promote interaction with the gas only. Many metal alkyls strongly absorb photons with wavelengths below 250 nm, allowing this method to be carried out with ArF ($\lambda$ = 193 nm) excimer lasers.[65,66,67,68] Photons from KrF (248 nm) excimer lasers and low-pressure Hg mercury lamps ($\lambda$ = 250 nm) can also be absorbed, although with a much lower cross section. A single photon can initiate cleavage of a valence bond in the molecule, resulting in multi-step dissociation of the excited radicals.[68] The particular reactions that take place depend on the specific molecule as well as the photon energy (owing to the molecules specific absorption spectrum). Various byproducts are generated in the process.[66] This technique has been used to grow CdTe[69,70,71] and CdHgTe[69,72] at temperatures as low at 165 °C with a high degree of uniformity. Likewise, single crystalline Si could be grown at substrate temperatures as low as 330 °C.[73]

One point that is worth mentioning is that irradiation of the gas phase has the ability to produce particles in the gas through parasitic reactions of the dissociated molecules. This was observed for the growth of $Cd_xHg_{1-x}Te$ using a $H_2$ carrier gas and a high pressure mercury lamp[69]. Comparatively, no dust formation was observed for CdTe growth using a low pressure mercury lamp.[70] It therefore appears that careful consideration of the precursor molecules, the carrier gas, and the photon spectrum and intensity could all be important factors in this process. Care must be taken to prevent such premature reactions.

Direct photon irradiation of the growth surface can also aid dissociation of adsorbed molecules. Two processes may operate in this regime: photolysis and catalysis. In the case of photolysis, molecules adsorbed on the substrate surface directly absorb photons and dissociate in a similar manner to those in the gas phase. This mechanism is mostly noted in the growth of III-V and Si epilayers. Experimental evidence shows that adsorption extends the photon absorption window to longer wavelengths, allowing KrF excimer laser and Hg lamp sources to be used to effectively dissociate precursors (i.e. trimethylgallium and silane) and drive growth even though those wavelengths are not efficiently absorbed in the gas phase.[74,75] GaAs growth using triethlygallium (TEG) precursors has even been demonstrated at 425 °C under photon irradiation from an Ar ion laser ($\lambda$ = 514 nm) suggesting that the absorption tail extends out to wavelengths beyond 500 nm.[76]

Sugiura *et al.* distinctly identified the mechanism for the growth of GaAs by metaorganic MBE under photon irradiation from an Ar ion laser was photolysis rather than catalysis by comparing the growth of GaAs and GaP.[76] Despite the difference in absorption of the laser light by the two different substrates (direct bandgap GaAs is highly absorbing, while indirect bandgap GaP is not), the growth rates were nearly identical, as shown in Fig. 5a. This result suggests that catalysis by photogenerated carriers was not the operative mechanism in the dissociation of the TEG precursors used in both growths. The build-up of reaction by-products on quartz reactor windows and other non-absorbing surfaces in other studies also supports this conclusion.[74,76,77] The growth rate of InP, also shown in Fig. 5, behaves in a similar manner.[78] The exact reaction pathway may be dependent on whether the molecule is physisorbed or chemisorbed.[79] Evidence for both single and multi-photon processes have been reported, depending on the chemistry.[68,79,80,81]

While the growth rate of II-VI semiconductors deposited at low temperatures by CVD exhibits a similar response to illumination conditions as III-V semiconductors (see Fig. 5b for the example of ZnSe), experiments have shown that photon-mediate growth changes are driven by catalysis.[74,82] The first substantial piece of evidence is the dependence of the growth rate on the photon energy, shown in Fig. 5c. For Zn-based chalcogenides, the growth rate substantially increases when the photon energy is above the film bandgap energy.[23,83,84,85] The second piece of evidence that catalysis is the primary factor in photon-assisted growth of II-VI semiconductors is that the quantum efficiency (number of dissociated molecules/number of incident photons) is ~10%.[23] This percentage is orders of magnitude higher than the efficiencies found for photo-assisted growth of other materials. Photogenerated carriers can therefore participate in the molecular dissociation process, perhaps through the elimination of an ethyl-metal bond by a hole.[82,83,86] Again, improved molecular dissociation rates can lead to an increase in the growth rate, as observed by several groups.[82,83,86,87] In many cases, improved surface morphology was also observed in cases where growth was intentionally carried out under kinetically limited conditions.[74,88]

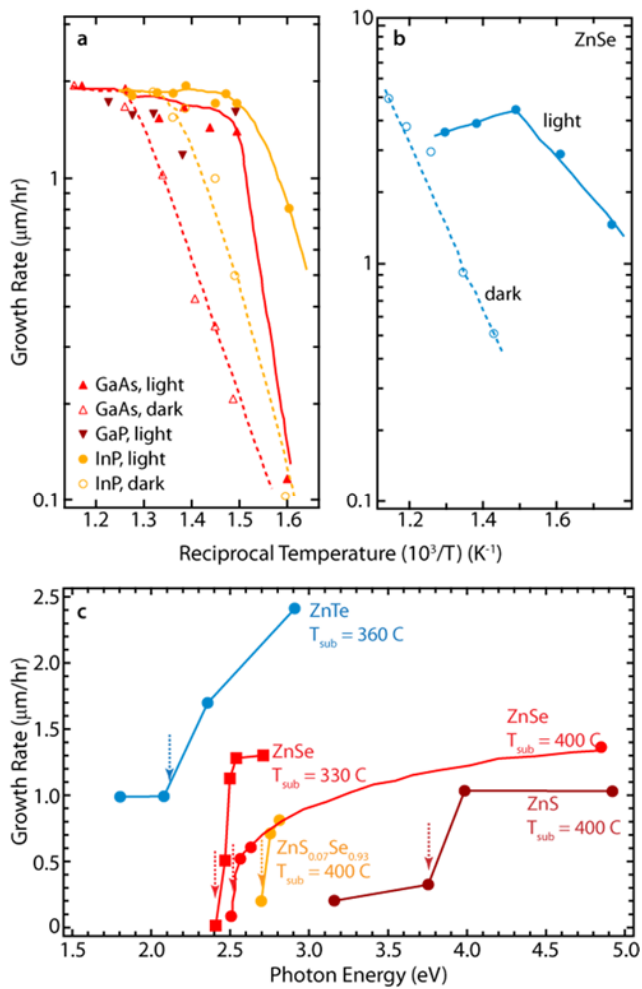

**Fig. 5** Growth Rates of III-V and II-VI semiconductors grown with metalorganic precursors. Growth rates of (a) GaAs, GaP and InP and (b) ZnSe as a function of substrate temperature and illumination conditions. (c) Growth rates of II-VI semiconductors grown under illumination as a function of photon wavelength. The bandgap energy of each semiconductor at the growth temperature is also marked with an arrow. Data was taken from Refs. 23,76,78,82 and 83.

Given that several growth processes may be operative during OMVPE, and that those processes are highly dependent on the specific growth conditions (i.e. temperature, precursor molecule, gas flow, substrate surface), photon irradiation may affect OMVPE growth differently based on variations in those conditions. This may account for the seemingly conflicting conclusions drawn from different studies. Specifically, molecular reactions in addition to processes involving single adatoms can be altered by light in growth techniques utilizing molecular precursors, making it difficult to definitively point toward a single mechanism.

A major advantage of the growth rate enhancement is the ability to grow epilayers in selective regions on the substrate because it can allow one to decouple the substrate temperature from precursor dissociation. Lowering the substrate temperature below that needed for thermal decomposition prevents growth from proceeding at substantial rates everywhere. Instead, light can be directed onto the areas of the substrate where growth is desired. Photochemical reactions can then be used to drive growth forward. The objective is to achieve a large differential in the growth rate between illuminated and non-illuminated regions.

Wankerl and co-authors demonstrated selective area growth of $Al_xGa_{1-x}As$ mesas from trimethylgallium (TMG), trimethylaluminum (TMA) and arsine gases at 500 °C using a tunable, frequency doubled dye laser.[89] They achieved a growth rate differential between dark and illuminated growth regions of 1:3.2, averaging a growth rate of 2.6 μm in the illuminated area over the first 15 minutes of growth. The growth rate differential depended strongly on the excitation wavelength, with selectivity dropping to near zero abruptly at wavelengths above 250 nm. The strong wavelength dependence indicates that the growth enhancement occurs through a photolytic process, where reaction of chemisorbed species is driven via direct light absorption. This hypothesis was supported by the fact that the alloy composition was the same between the regions grown under illumination or in the dark, indicating that the growth mechanism did not lead to different reaction rates between TMG and TMA. Photolysis (rather than pyrolysis, a reaction driven by photogenerated carriers or photodissociation of molecules in the gas phase) is the preferred process for selective area growth because it offers the highest spatial resolution. Other processes involving temperature, free carriers or reacted gas molecules allow for diffusion over relatively wide length scales, reducing the ultimate resolution of the process.

*Material Quality*

An enticing observation of photo-assisted growth has been an improvement in material quality. These improvements were most significant when the growth was carried out at lower-than-normal substrate temperatures. Growth at reduced temperatures is generally advantageous for improving dopant incorporation and limiting atomic diffusion between epilayers. However, optical and transport properties as well as surface smoothness tend to deteriorate. Thus, a trade-off is usually sought. Photo irradiation may offer the best aspects of both approaches. Photo-assisted MBE was shown to substantially improve the low temperature free exciton emission from ZnSe epilayers grown at substrate temperatures as low as 150 °C to intensities emitted from epilayers grown at more optimal temperatures around 250-300 °C.[9] Similar changes in low temperature photoluminescence (PL) associated with improved crystallinity were demonstrated for ZnS.[10] Growth of polar III-V semiconductors on non-polar Si substrates is also known to reduce crystalline quality by introducing extended antiphase domains due to random formation of group III and V sublattices. Photo-assisted epitaxy of GaAs on Si using a pulsed KrF excimer laser has been shown to improve the optical properties of the GaAs epilayer.[90]

Many of the studies mentioned above explicitly point out that the improvements in material quality observed under photon irradiation were not the result of local heating and instead may be attributed to electronic effects. One such possibility is that carrier-assisted anion desorption leaves a metal-rich growth surface, which promotes an increase in adatom mobility and a general improvement in the crystal structure (i.e. atoms residing on energetically favorable sites).[61] A transition to metal rich surfaces was experimentally verified by Wu on static and dynamic CdTe growth surfaces.[91] Irradiation of the growth surface with an Ar ion laser (500 mW/cm$^2$) caused the surface reconstruction to shift from a Te-stabilized (2x1) to a Cd-stabilized c(2x2) pattern, indicating that Te had preferentially desorbed. Since cations tend to stick to anions but not to other cations, a cation metal rich surface allows those atoms to diffuse over longer distances on the growth surface before finding favorable incorporation sites. Enhanced adatom mobility and longer diffusion lengths therefore allow growth to proceed via a 2D growth mode rather than a 3D mode. This effect is especially influential at low growth temperatures, where adatom diffusion lengths are generally lower and 3D growth tends to dominate. Indeed, the improvements in the material quality of ZnS epilayer grown by MBE at low temperature when subjected to photon irradiation were linked to a change from 3D growth to 2D growth through RHEED measurements.[10]

Improvement in "material quality" is a fairly general assessment. Point and extended defects, grain boundaries and orientation, and surface morphology all play a role in the structural, optical and electronic properties of semiconductor epilayers. The influence of photon irradiation on each of these aspects is detailed in the following sections.

*Point Defects*

Native defects have a substantial effect on the optical and transport properties of semiconductor epilayers and devices, impacting the overall "material quality" assessment of the material. Photo-assisted growth can affect their concentration through photogeneration of non-equilibrium carrier populations. Addition of a non-zero charge to a defect requires the real exchange of carriers between the defect and either the conduction or valence band. The influence of non-equilibrium carrier concentrations on defect formation was originally viewed in the framework of the mass-action law, wherein the distribution of a defect in different charge states is determined in part by the availability of free carriers with which it can react.[61,92] For example, defect $X$ could undergo the reaction $X^0 + e^- \rightarrow X^{-1}$. Simply put, additional carriers drive the reaction forward. Calculations based on this approach were carried out by Ichimura, *et al.*, for the case of p-type ZnS and ZnSe.[92] Their results indicated that the concentrations of positively and negatively charged vacancies decreased as the concentration of non-equilibrium electrons increased.

More recently, this problem has been re-framed in the context of the Fermi level. First principles calculations of defect formation energies and defect concentrations account for the carrier exchange mechanism by including a Fermi level term, $\mu_F$, in the defect formation enthalpy, $H^q_f$. For compound AB, $H^q_f$ can be written as:

$$\Delta H^q_f = E_D + q(\mu_F + E_{VB}) - n_A \mu_A - n_B \mu_B \qquad (2)$$

where $q$ is the charge state, $E_{VB}$ is the position of the valence band maximum, $E_D$ is the total energy of the supercell comprised of $n_A$ and $n_B$ atoms and a single charged defect, and $\mu_A$ and $\mu_B$ are the chemical potentials of atoms $A$ and $B$. $\mu_A$ and $\mu_B$ represent the partial pressures of $A$ and $B$ species and their internal degrees of freedom as they arrive at the substrate surface. Growth of the AB compound under thermodynamic equilibrium is typically carried out under the control of three parameters: the temperature, $\mu_A$ and $\mu_B$. The Fermi level is then self-consistently determined by the concentration of native defects and extrinsic dopants under the condition of charge neutrality.

The addition of non-equilibrium carrier concentrations to the semiconductor through light exposure splits the electron and hole quasi Fermi levels (QFLs), $\mu_{Fn}$ and $\mu_{Fp}$, changing the chemical potential of each. Carrier exchange between the defects and conduction and valence bands can proceed at different rates than under dark growth conditions. Splitting the QFLs therefore adds an extra degree of freedom that can be manipulated during growth. A simplified model to account for the effect of QFL splitting on the defect formation energy was first presented by Bryan *et al.*, where the overall Fermi level could be written as an average of $\mu_{Fn}$ and $\mu_{Fp}$.[14] Subsequently, this model was revised to include a more thorough accounting of the relative contributions of the QFLs based on the electron and hole capture and emission rates that control carrier exchange.[50,93] The Fermi level contribution in Eqn (2) can be modified as:

$$\mu_F = \left( \frac{r_{c,n} + r_{e,n}}{r_{c,n} + r_{e,n} + r_{c,p} + r_{e,p}} \right) \mu_{Fn} + \left( \frac{r_{c,p} + r_{e,p}}{r_{c,n} + r_{e,n} + r_{c,p} + r_{e,p}} \right) \mu_{Fp} \qquad (3)$$

where $r_{c,n}$ is the electron capture rate, $r_{e,n}$ is the electron emission rate, $r_{c,p}$ is the hole capture rate, and $r_{e,p}$ is the hole emission rate (see Fig. 6). Equation 3 reverts back to $\mu_F$ under thermodynamic equilibrium (i.e. absence of photogenerated carriers). It is also equivalent to the original formulation set forth by Bryan *et al.* in Ref. 14 under the special circumstances where the charge transition level of the defect is equal to the intrinsic Fermi level and the carrier capture cross sections, thermal velocities and effective densities of states for the conduction and valence bands are equal.[93]

The dominant carrier capture or emission rate will determine which QFL most influences the defect formation energy in a particular system. When the $\mu_F$ term in Eqn. 3 is dominated by the minority carrier QFL of the opposite charge of the defect (i.e. for compensating defects in a doped semiconductor), the defect concentration is generally predicted to decrease. However, when the $\mu_F$ term is dominated by the minority carrier QFL of the same charge as the defect, the defect concentration may actually increase. The later behavior is more likely to occur in wide bandgap semiconductors, where the charge transition levels of deep defect states are located far from either the conduction or valence band edge. This has the effect of suppressing carrier emission to that band, allowing the capture rate to automatically dominate.

Of course, this assessment assumes that the photocarrier generation rate is high enough to move the dominant QFL. In cases where this condition is not met (for example, the photocarrier generation rate is low, the semiconductor is highly doped and the $\mu_F$ term is dominated by the majority carrier QFL) the defect concentration is not expected to change. A more detailed analysis of possible behaviors is presented in Ref. 93 for a generic deep donor defect. Quasi-Fermi level splitting will obviously have a large impact on compensating defects in doped semiconductors and can affect the overall free carrier concentration. A more thorough accounting of this aspect is provided below in the *Increased Doping Efficiency* section.

It should be noted that this theoretical analysis is only appropriate for growth that occurs under otherwise thermodynamic limitations. Defects may fail to form or may form at higher concentrations if growth processes are substantially kinetically limited.

*Extended Defects*

Light irradiation during growth has also led to a reduction of extended defects in heterostructures. One example is the MBE growth of HgCdTe on lattice-mismatched (001) CdZnTe substrates at 200 C.[13] Under photon irradiation by a 514 nm Ar+ laser line (40 mW/cm$^2$), these films exhibited lower and more reproducible dislocation densities ($\sim 1 \times 10^5$ cm$^{-2}$) compared to films grown by conventional MBE ($10^5 - 10^6$ cm$^{-2}$). Microscopy revealed longer and more highly aligned misfit dislocations in the samples grown under photon irradiation, which would allow associated threading dislocations to more easily annihilate. Other groups reported similar improvements in HgCdTe epilayers grown on lattice-matched CdZnTe substrates at even lower temperatures of 180 C.[94] Threading dislocation densities as low as $3 \times 10^4$ cm$^{-2}$ were measured in some samples, while the double-crystal rocking curve full width at half maximum (FWHM) was reduced to values (25 – 30 arc-s) typically observed in homoepitaxial GaAs. The changes in the extended defect structure led to improvements in the transport properties of both n and p-type HgCdTe epilayers.[94]

No specific mechanisms responsible for the observed extended defect modifications were identified in the case of the HgCdTe growth studies, which were all carried out at temperatures ≤ 200 °C. However, photo-assisted growth could be used to suppress dislocation formation via strain relaxation by permitting a reduction in the growth temperature, removing some of the thermal energy needed for dislocation formation. This was demonstrated in the growth of single CuIn$_x$Ga$_{1-x}$Se (CIGS) on GaAs substrates, where there is a lattice misfit of ~2.3%.[95] Molecular beam epitaxy of CIGS is usually carried out around 475 °C, but the use of UV photon irradiation allowed the substrate temperature to be dropped to 300 °C without significant breakdown in the crystalline structure. As a result, strain was only partially relieved through microtwin formation rather than full relaxation through dislocation generation. Strong suppression of strain relaxation has also been observed when semiconductors are grown at low temperatures without photon irradiation.[96]

*Surface Roughness*

Surface roughness is principally controlled by the growth mode. Light stimulation has the potential to affect a number of processes that ultimately control the growth mode, including desorption, adatom mobility and the availability of preferential bonding sites. Photo-assisted growth can therefore affect the surface roughness through many mechanisms and indeed appears to have a range of effects.

Reports of improvements in surface smoothness under light stimulation typically occur when growth is carried out at relatively low temperatures. This includes the epitaxy of ZnS at 150 °C and growth of GaN on SiC substrates at temperatures near 600 °C (where growth is typically carried out at substrate temperatures in excess of 800 °C).[10,97,98] Enhanced smoothness may be linked to the preferential desorption of anion species, as discussed above. Increased adatom mobility is likely to promote 2D growth under conditions where 3D growth usually dominates (i.e low temperature growth).

There are also reports of increased surface roughness as the result of photon irradiation, especially when growth is carried out near optimal substrate temperatures for dark growth conditions. These include the growth of ZnSe on GaAs substrates at temperatures in the range of 250-400 °C as well as homoepitaxy of GaAs at substrate temperatures 550 – 650 °C.[60,99,100] In certain cases, the observation of increased surface roughness is linked to the promotion of a 3D growth mode. For example, photon irradiation from a Xe lamp during GaAs MBE on 2° miscut GaAs substrates induced a 20 °C increase in the temperature at which layer-by-layer growth transitioned to step-flow growth.[100] At first glance, it appears possible that such an increase may result from a reduction in the adatom mobility on the surface, requiring higher temperatures for the adatoms to travel the same distance to the step edges. However, a similar effect was observed for GaAs MBE in the presence of hydrogen.[101] In that case, it was suggested that hydrogen passivated dangling bonds at the step edges, reducing their potential as preferential bonding sites. The energetic cost for Ga adatoms to attach to step edges became similar to the cost of forming new nucleation sites on the terraces. This situation may also be the cause of the transition from step-flow to layer-by-layer growth in the case of light-stimulated epitaxy, where photogenerated carriers could provide a similar passivation effect.[100]

One interesting item of note related to adatom diffusion is that photon irradiation appears to, in some cases, affect the diffusion of impurities on semiconductor surfaces differently depending on whether they are an n-type or p-type dopant.[38,102] This was observed for the diffusion of In (p-type) or Sb (n-type) atoms on Si (111) surfaces. The diffusivity of Sb increased under illumination, while that of In decreased. No definitive mechanism has been experimentally identified. However, possible sources have been linked to a change in the formation of impurity-vacancy complexes, which could help to drive diffusion. These complexes may be more or less likely to form based on the conditions of Fermi level pinning at the Si surface, the direction and magnitude of the associated band bending, and QFL splitting. More information is needed to understand the origins of this effect, but based on the greater picture of changes to defect populations, it is likely to be electronic in nature.

*Grain Orientation*

In addition to influencing extended defect formation, UV irradiation has been found to change the texture of polycrystalline semiconductors grown on a variety of substrates. This includes heteroepitaxial growth of GaN on β-SiC,[98] ZnS on quartz and sapphire,[103,104] as well as Si homoepitaxy.[8] In all cases, the grains became more highly oriented under irradiation with photons of above-bandgap energies. Careful investigation by Yokoyama, *et al.*, into the growth of ZnS on sapphire showed that changes in grain orientation were most pronounced when irradiation was applied either during the initial stages of film growth or throughout the entire deposition.[104] Little to no effect was observed when irradiation was applied only at the end of the deposition or only before deposition had started. These results suggest that the grain orientation is largely set during the initial states of film growth and not due to changes to the starting substrate surface. Alteration of the adatom mobility or arrangement on the surface was cited as general reasons for the preferential texturing, but no definitive mechanisms were uncovered.[98,104] However, based on other studies of photo-assisted growth, the interaction of photogenerated carriers with dangling bonds at the surface may help to promote or block adatom incorporation along certain crystalline directions.[100]

*Increased Doping Efficiency*

Another exciting observation associated with photo-assisted epitaxy has been the increase in free electron and hole concentrations, especially in II-VI semiconductors. Since low doping efficiencies have generally limited the performance of II-VI semiconductors in optoelectronic devices, including solar cells, LEDs and lasers, this was a particularly promising development.[105] Room temperature free electron concentrations as high as $8 \times 10^{17}$ cm$^{-3}$ and hole concentrations as high as $6 \times 10^{18}$ cm$^{-3}$ have been measured in In-doped and As-doped CdTe, respectively.[12,106] Both results were obtained by photo-assisted MBE growth, and films grown under the same conditions without photon irradiation were generally insulating. Enhanced free carrier concentrations were also observed in Al-doped ZnS grown by e-beam deposition[103] and N-doped ZnSe grown by both OMVPE and MBE.[107] More recently, elevated free electron concentrations were observed in Al$_x$Ga$_{1-x}$N alloys grown by photo-assisted OMVPE, extending the improvements to III-V semiconductors as well.[14]

Several mechanisms have been invoked to explain these results. One early explanation was that enhanced anion desorption under photon irradiation creates openings in the host atomic reconstruction on the epilayer surface that makes it easier for dopant atom incorporation.[11,108] So far, this argument has only been supported by indirect experimental evidence. For example, preferential desorption of Te from a CdTe surface under UV irradiation compared to Cd or dopants such as Sb was observed through changes in static RHEED reconstruction patterns or Auger spectra.[11] Presumably, this indicated that Sb was more likely to incorporate into the epilayer. In ZnSe, a reduction in the growth rate under illumination due to excessive Se desorption combined with an increase in the acceptor density, measured by electrochemical profiling, was concluded to be caused specifically by an increase in N acceptors.[108] It should be noted that the N concentration in these films was not directly measured by techniques such as SIMS in this study. Despite the attractiveness of this argument for explaining the improved dopant incorporation on the anion site, it is less likely to apply to improvements in the incorporation of dopants that substitute on the cation site, since the cations are much more stable to desorption under light. It was suggested that enhanced Te desorption could prevent the formation of In$_2$Te$_3$ secondary phases that would reduce the concentration of substitutional In$_{Cd}$ donors in order to explain the improvement in the free electron concentration in In-doped CdTe, but no experimental evidence has been provided.

The other primary explanation is that photon irradiation helps to reduce native defects, which would otherwise compensate the extrinsic dopant atoms. Since the formation energy of charged defects changes linearly with the Fermi level, it can vary substantially in wide bandgap semiconductors. Moving the Fermi level toward either the conduction or valence band through the addition of extrinsic dopants typically results in very low formation energies for compensating defects, sometimes on the order of the extrinsic dopant density itself.[109,110] As described above, excess photogenerated carriers have the potential to alter the contribution of the Fermi level to the defect formation energy, leading in some cases to a reduction in compensating defect concentrations. Experimental evidence of reduced compensating defect concentrations produced under photo-assisted growth primarily comes from low temperature PL. Such measurements are particularly useful for studying changes in defects in that they are highly sensitive to radiative recombination from shallow donor and acceptor states. Giles, *et al.*, showed that the emission from In-doped CdTe epilayers grown without photon irradiation was dominated by an array of vacancy-donor and other defect-related transitions at 1.45 eV and shallower donor-acceptor pair (DAP) transitions around 1.55 eV (see Fig. 7).[111] Those transitions were strongly suppressed in similar In-doped epilayers grown under illumination from an Ar ion laser. They were replaced by donor-related transitions around 1.58 eV. The absence of these transitions in similarly grown CdTe epilayers without In doping suggests that those donor transitions were related to substitutional $In_{Cd}$ donors (see Fig. 7). The disappearance of the deep acceptor-related emission peaks when CdTe:In was grown under above-bandgap photon irradiation indicates a sharp reduction in charged compensating defects. In wide bandgap nitrides, a relative drop in the intensity of deep vacancy-related PL from Si-doped AlGaN epilayers was observed when photon irradiation was used during OMVPE growth.[14] Reduction in donor-acceptor luminescence was also observed from GaN epilayers subjected to intense electron irradiation during MBE growth.[112]

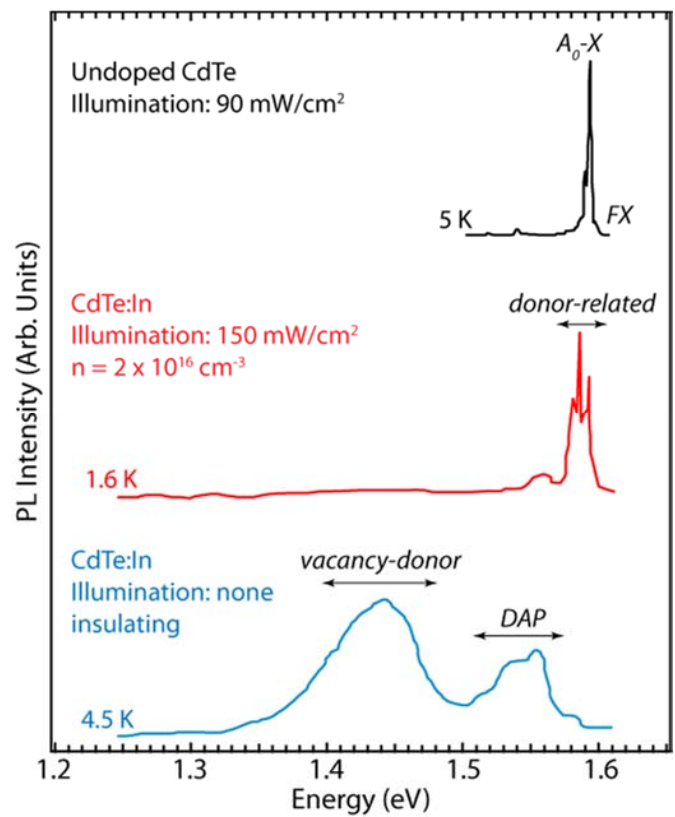

**Fig. 7** Low temperature PL spectra CdTe with and without In doping, grown under a range of illumination conditions. Data was taken from reference 111.

In reality, a balance of desorption and compensating defect reduction mechanisms likely contributes to enhanced extrinsic doping efficiencies, depending on the particular semiconductor and dopant as well as the exact growth conditions. A study by Simpson *et al.*, suggested that N incorporation into ZnSe increases with lower Se fluxes and/or Se desorption under irradiation from a Kr ion laser, possibly contributing to increased acceptor concentrations, as noted above.[108] However, PL measurements indicated a change in deep donor concentrations, possibly associated with $V_{Se}$-Zn-$N_{Se}$ defects in samples grown under illumination. The theoretical framework for how the Fermi level contribution changes in the presence of non-equilibrium carrier concentrations is based on thermodynamic constraints. MBE and OMVPE growth, especially of II-VI semiconductors containing elements with high vapor pressures, are typically carried out at low temperatures in kinetically-limited regimes. The exact types and concentrations of defects that are formed will likely depend on a host of factors, including the substrate temperature and material, surface morphology and the particular dopant species that is incorporated. In the case of OMVPE growth, molecular reaction dynamics could also play a significant role. Certainly, more investigation is needed to fully understand the different effects that photon irradiation can have on extrinsic doping. A large body of experimental evidence already points towards improved doping efficiency with photo-assisted growth. However, very few studied have systematically focused on the effects of photon irradiation conditions, substrate temperature and dopant flux. This information is critical for confirming the mechanisms responsible for doping improvement.

*Incorporation*

Extrinsic dopant atoms are also prone to diffusion from an epilayer of high doping concentration to a subsequently-grown epilayer of lesser doping in a process known as dopant carry-forward.[113] This phenomenon is particularly problematic because the dopant atoms can trail into the adjacent low-doped layer over tens of nanometers, effectively limiting the ability to create abrupt doping profiles. The root cause of this effect appears to be two-fold. Fermi-level pinning near mid-bandgap at the growth interface of the doped layer sets up band bending and an electric field over a few nanometers into the epilayer.[113,114] Native defects such as vacancies also preferentially form under high doping conditions, as described above.[115,116,117] Both the electric field and presence of defects can aid extrinsic dopant atoms in continuously diffusing to the growth surface, where they amass and slowly incorporate as the growth proceeds. Shown in Fig. 8, light stimulation during the transition in growth of a highly Si-doped GaAs epilayer to an undoped GaAs epilayer by MBE was shown to reduce carry-forward of the Si dopants from 76 nm/decade in the dark to 23 nm/decade under excitation from a KrF excimer laser.[118] While the relative effect of light stimulation on the two leading mechanisms of carry-forward was not distinguished in this initial demonstration, both are expected to be affected.

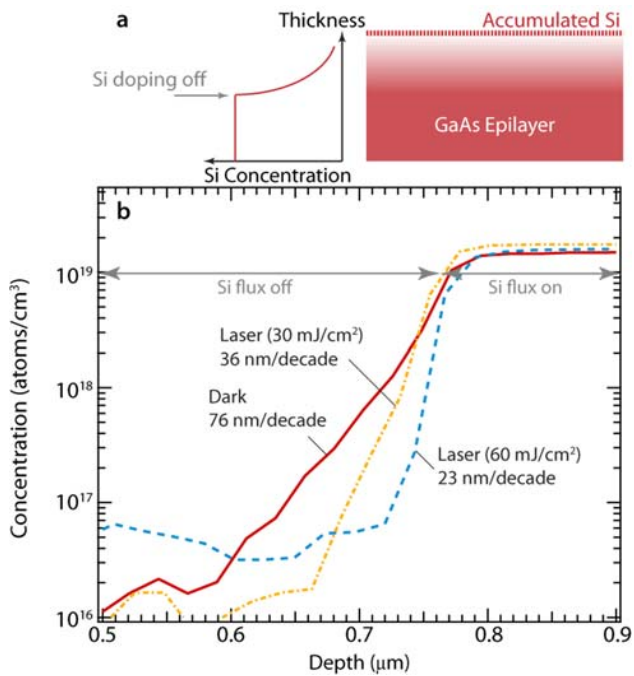

**Fig. 8** Dopant carry-forward. (a) Schematic of the dopant carry-forward process for Si dopants in GaAs. (b) Concentration of Si in GaAs as a function of depth and illumination conditions. The depth at which the Si flux was turned off is marked with arrows. Data is from Ref. 113.

Finally, light has been observed to affect the incorporation of isoelectronic Bi atoms into GaAs by MBE.[16] The large atomic radius of Bi causes it to act as a surfactant, and incorporation is only achieved at low temperatures < 400 °C. Low temperature PL spectra of $GaAs_{1-x}Bi_x$ grown under dark conditions with less than 1% Bi include emission from bound Bi pair states located just above the valence band, as shown in Fig. 9. However, samples with similar Bi concentrations grown under UV illumination from a KrF excimer laser reveal a lower concentration of Bi pair states. A change in the Bi pair density for the same alloy composition indicates two things: 1) Bi pairs may form at higher densities under dark conditions than would occur through random incorporation, and 2) light appears to inhibit their preferential formation. The mechanism for Bi pair suppression is not currently understood. One possibility is that Bi arrives at the growth surface primarily as dimers (just as $As_2$ or $P_2$ anions do), and those atoms tend to go into the lattice near each other. Photogenerated carriers may drive the dissolution of those dimers at a higher rate, allowing them to move apart before they incorporate. Another possibility is that enhanced desorption of As atoms from the surface reconstruction opens up additional individual spaces for Bi atoms to incorporate, again forcing Bi atoms to incorporate separately. More information about this process is needed.

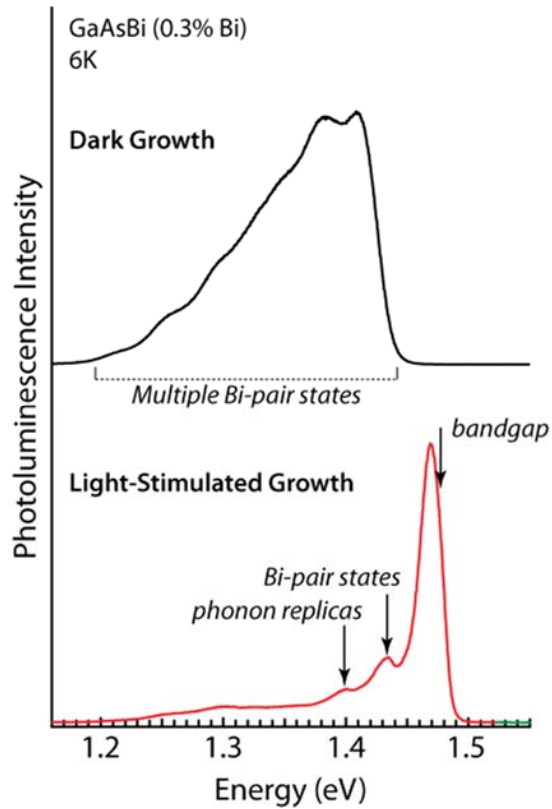

**Fig. 9** Low temperature PL spectra of GaAsBi with 0.3% Bi grown under (top) dark and (bottom) light-stimulated growth using a KrF excimer laser. Data was taken from Ref. 16.

**Considerations and Future Directions**

*Summary of Effects*

The investigations and results discussed above indicate that when light is applied to the growth surface judiciously, it alters semiconductor epitaxy primarily through electronic rather than thermal means. In the vast majority of cases, the photon generates free carriers that then go on to impact the electron and hole potentials, desorption rates and bonding mechanisms. Changes in these mechanisms then modify a number of growth processes, including defect formation, elemental segregation, incorporation, adatom mobility, morphology and chemical reactions. Adatom mobility appears to be regulated by both desorption and bonding mechanisms, making it difficult to separate which ones are responsible for changes observed in epilayer growth. In a smaller percentage of cases, photons can be absorbed by molecules directly and participate in molecular dissociation or other chemical reactions. Figure 10 presents an overview of the relationship between these mechanisms and the end growth processes.

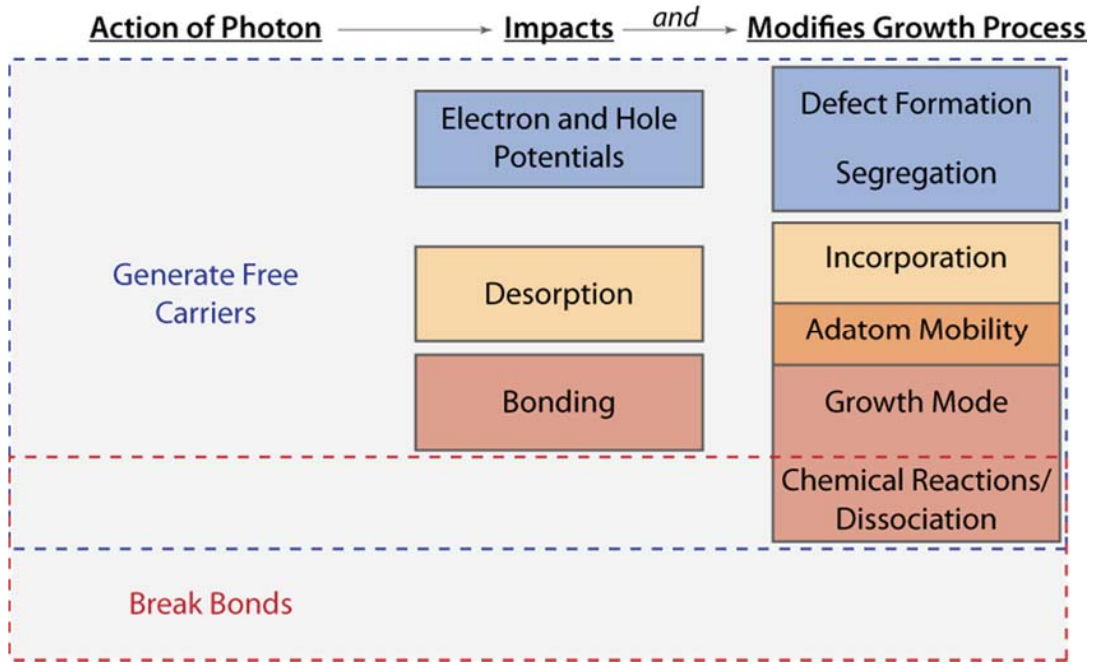

**Fig. 10** Growth processes that are impacted by photon irradiation of the growth surface.

Enhanced desorption is observed much more predominantly for anion species than cations. It can change the competition for anion sites within the lattice, whether that is between isoelectronic anion species in an alloy or between the host anion and a dopant. A change in the site competition dynamics may extend solubility limits or doping efficiencies. Decreased anion coverage can also leave the surface metal-rich, which has been linked to improved surface morphology and has been offered as a potential explanation for improving the material quality of low temperature-grown epilayers. Quasi-Fermi level splitting through the optical injection of excess free electrons and holes can profoundly modify defect formation energies and built-in electric fields at the growth surface. This effect has led to changes in the solubility and diffusion of charged ions and the abruptness of doping profiles and heterovalent heterostructures. Passivation of dangling bonds on the surface can also alter the number of favorable incorporation sites and even alter the growth mode.  The fact that photogenerated carriers are the root source of these modifications means that their density must be greater than the intrinsic carrier concentration in the semiconductor to create a measureable excess that will precipitate these changes.

Collectively, the results discussed above provide some insight into the ways in which photo-assisted growth techniques can be applied favorably. First, light appears to produce the greatest changes in growth processes and the resulting material properties when it is applied to growth at low-temperatures.  This is likely due to the fact that 1) low temperature growth typically results in poorer material quality to start with, and 2) the intrinsic carrier concentration that must be overcome with photogenerated carriers is smaller at lower temperatures. Clearly, the ability to maintain high crystal quality at low growth temperatures has a number of benefits. Low temperatures may be selected to reduce interlayer atomic diffusion, suppress strain relaxation and limit the thermodynamic driving force for defect formation.  This last point may prove to be important in improving the doping efficiency or overall free carrier concentration in doped semiconductors.  Second, light can be used to intentionally de-couple processes that are affected by temperature. This includes desorption, adatom diffusion, growth modes and chemical reactions. For example, the substrate temperature can be held relatively low, but light can be used to selectively drive certain chemical reactions. Photon irradiation can essentially act as an additional controllable growth parameter.

*Outlook and Future Areas of Exploration*

The view of light as an additional growth parameter naturally leads to the question of whether variation in other parameters might result in the same effects. For instance, could lower anion fluxes be used instead of enhancing anion desorption with light? The prospect of creating the same conditions without light might be possible in some instances. Yet, modification of the electron and hole QFLs is not possible with temperature and flux alone.

There are several areas where we still lack a proper understanding of the mechanisms by which photons can affect growth processes. We suggest the following as topics of future research and also outline some approaches for extracting meaningful information.

- <u>Enhanced Doping Efficiency</u>: It is still not clear whether the improvement in free carrier concentration that has been achieved under photo-assisted growth is primarily due to an increase in dopant atoms or a decrease in compensating defects. This is due in part to the lack of a systematic study that specifically measures differences in the concentrations of each as a function of irradiation conditions. Rather, many of the previous studies investigated only a few samples and inferred changes in the dopant and defect populations through indirect means, such as PL intensities or total acceptor density. Instead, these changes should be determined through a combination of direct quantitative measurements, including secondary mass ion spectrometry (SIMS) to measure the total concentration of dopant atoms and deep level transient spectroscopy (DLTS) to measure the concentration and energy of different trap states. These measurements should be carried out on a comprehensive set of samples that were grown under a range of irradiation conditions with both n and p-type doping and perhaps different doping species that reside on anion and cation sites. As mentioned above, both mechanisms may be operable at the same time, depending on the specific growth conditions used. If possible, multiple material systems should be studied in order to observe general trends.

- Modification of Native Defect Formation Under Light Stimulation: This topic is linked to the topic of enhanced doping efficiency. In particular, we still need to experimentally verify the theoretical framework laid out above that describes how QFL splitting may affect the defect formation energy. Native defect concentrations should be quantified as a function of growth and illumination conditions. Material systems with well-known native defect types and formation energies should also be chosen to aid in their identification experimentally. One considerable complication is that the theoretical treatment of this problem was carried out under thermodynamic constraints, but growth is often kinetically-limited to some degree (especially low temperature-grown II-VI semiconductors). We therefore suggest that a more appropriate way to test the relationship between QFL splitting and defect formation is to probe the change in defect concentrations in semiconductor crystals after annealing for long times under light. Such an environment could be made to be as close to thermodynamically-limited as possible. Again, the effect should be tested for both n and p-type doping to modify the degree to which each QFL moves under illumination.

- Adatom Movement on the Growth Surface: Since improvements in the material quality of epilayers grown at low temperature are linked to enhanced adatom mobility, it is important to understand the origin(s) of the enhancement. This understanding is expected to impact areas beyond epitaxy, including diffusion of atoms on semiconductor or metal surfaces in general. From observations in the literature and the processes compiled in Fig. 10, increased desorption and/or changes in the bonding behavior of adatoms may contribute to this phenomena. Those mechanisms should be distinguished.

- Incorporation Dynamics: The ability to alter how atoms are incorporated into an epilayer could be very powerful. Whether it is increasing uniformity on a macroscale or changing the distribution of elements in anion dimers, light appears to be a tool that will help us to tailor the composition, structure and properties of semiconductor epilayers and heterostructures. There are still many things to learn. The exact mechanisms behind modifications to compositional segregation and anion intermixing are not definitively known.

Lastly, photo-assisted epitaxy techniques also have the potential to play an important role as we continue to push the boundaries of semiconductor epitaxy. Such endeavors include low temperature synthesis of metastable materials or heteroepitaxial growth of 2D and bulk epilayers on substrates with dissimilar lattice constants, chemical properties or even crystal structures. The ability to access lower growth temperatures without sacrificing adatom mobility, chemical reaction rates or material quality will certainly be useful in this endeavor. One or more of the mechanisms discussed above may be exploited to control surface properties and growth processes. Advancing our understanding of how photons alter growth mechanisms will aid us in determining whether photo-assisted epitaxy techniques may be beneficial in individual circumstances.

**Acknowledgements**

K.A. acknowledges financial support from the Department of Energy Office of Science, Basic Energy Sciences under contract DE-AC36-08GO28308. M.A.S. acknowledges support from the Department of Energy through the Bay Area Photovoltaic Consortium award DE-EE0004946.

118 C.E. Sanders, D.A. Beaton, R.C. Reedy and K. Alberi, *Appl. Phys. Lett.*, **106**, 182105 (2015).